\documentstyle[12pt,psfig]{article}
\begin{document}
\begin{titlepage}
\begin{center}

{\Large\bf{The DLLA limit of BFKL in the Dipole Picture }}
\\[5.0ex]
{\Large\it{ M. B. Gay  Ducati $^{*}$\footnotetext{$^{*}$E-mail:gay@if.ufrgs.br}}}\\
 {\it and}\\
{ \Large \it{ V. P. Gon\c{c}alves $^{**}$\footnotetext{$^{**}$E-mail:barros@if.ufrgs.br} 
}} \\[1.5ex]
{\it Instituto de F\'{\i}sica, Univ. Federal do Rio Grande do Sul}\\
{\it Caixa Postal 15051, 91501-970 Porto Alegre, RS, BRAZIL}\\[5.0ex]
\end{center}

{\large \bf Abstract:}
In this work we obtain the 
DLLA limit of BFKL in the dipole picture and compare it  with  HERA data.
We demonstrate that in leading - logarithmic - approximation, where 
$\alpha_s$ is fixed, a transition  between the BFKL dynamics and the  
DLLA limit  can be obtained in the region of $Q^2 \approx \,150\,GeV^2$.
We compare this result with the  DLLA predictions obtained with 
$\alpha_s$ running. In this case a transition  is obtained at low $Q^2$
($ \le 5\,GeV^2$).  
This demonstrates the importance of the next-to-leading order corrections
to the BFKL dynamics.
Our conclusion is that the $F_2$ structure function is not the best 
observable for the determination of the dynamics, since there is great 
freedom in the choice of the parameters used in both  BFKL and DLLA predictions.

\vspace{1.5cm}



\end{titlepage}

\section{Introduction}

       The behaviour of $ep/pp$ scattering 
in  high energy limit  and fixed 
momentum transfer is one of the outstanding open questions in the theory of 
the strong interactions.  In the late 1970s, Lipatov and collaborators 
\cite{bfkl} established the papers which form the core of our knowledge of Regge limit (high energies limit) of Quantum Chromodynamics (QCD). The physical effect that they describe is often 
referred to as the QCD pomeron, or BFKL pomeron.
The simplest process where the BFKL pomeron applies is the very high energy 
scattering between two  heavy quark-antiquark states, i. e. the onium-onium 
scattering. For a suficiently heavy onium state, high energy scattering is a 
perturbative process since the onium radius gives the essential scale at 
which the running coupling is evaluated. Recently \cite{mueller, mueller1, 
mueller2} this process was studied in the QCD dipole picture, where the 
heavy quark-antiquark pair and the 
soft gluons in the large $N_c$ limit are viewed as a collection of color 
dipoles. In this case, the cross section can be understood as a product of 
the number of dipoles in one onium state times the number of dipoles in the 
other onium state times the  basic cross section for dipole-dipole 
scattering due to two-gluon exchange. In \cite{mueller} Mueller demonstrated that the 
QCD dipole picture reproduces the  BFKL physics. In this work we  discuss 
the BFKL pomeron using the QCD dipole picture.

       Experimental studies of the BFKL pomeron are at present carried out 
mainly in HERA {\it ep} collider in deeply inelastic scattering in the 
region of low values of the Bjorken variable $x \equiv \frac{Q^2}{2p.q}$, 
where $Q^2\equiv -q^2$. Here $p$ is the four-momentum of the proton and
$q$ is the four-momentum transfer of the eletron probe. 
In this case, QCD pomeron 
effects are expected to give rise to a power-law growth of the structure 
functions as $x$ goes to zero. However, this study in 
deeply inelastic scattering is made difficult by the fact that the low $x$ 
behaviour is influenced by both short distance and long distance physics 
\cite{bartels}. 
As a result, predictions at photon virtuality $Q$ depend on a   
nonperturbative input at a scale $Q_0 < Q$. This makes it difficult to 
desantangle perturbative BFKL predictions from nonperturbatives effects. 
Moreover, the program of calculating the  next-to-leading  corrections to 
the BFKL equation was only  formulated recently \cite{fadin}. Of course 
there are uncertainties due to subleading corrections and from the 
treatment of infrared region of the  BFKL equation which will modify the 
predictions of this approach.

        One of the most striking discoveries at  HERA is the steep rise 
of the proton structure function $F_2(x,Q^2)$ with decreasing Bjorken $x$ 
\cite{h1}.
The behaviour of structure function at small $x$ is driven by the gluon 
through the process $g\rightarrow q\overline{q}$. The behaviour of the 
gluon distribution at small $x$ is itself predicted from perturbative QCD
via BFKL equation. This predicts a characteristic $x^{-\lambda}$ singular 
behaviour in the small $x$ regime, where for fixed $\alpha_s$ the BFKL 
exponent $\lambda = \frac{3\alpha_s}{\pi}4ln2$. It is this increase in the 
gluon distribution with decreasing $x$ that produces the corresponding 
rise of the structure function. However, the  determination of the valid 
dynamics in the small $x$ region is an open question, since the conventional 
DGLAP approach \cite{dglap} can give an excellent description of $F_2$ at 
small $x$. The goal of this letter is  to discuss this question.

Recently Navelet {\it et al.} \cite{navelet} applied 
the QCD dipole model to deep inelastic lepton-nucleon scattering. They 
assumed that the virtual photon at high $Q^2$ can be described by an onium. 
For the target proton, they made an  assumption that it can 
be approximated by a collection of onia with an average onium 
radius to be determined from the data. This model described reasonably the 
$F_2$ data in a  large range of $Q^2 (< 150\, GeV^2)$ and 
 $x$. 

In this letter we obtain the double-leading-logarithmic-approximation 
(DLLA) limit  of BFKL in the QCD dipole picture using the approach proposed 
by Navelet {\it et al.}. This limit is common to BFKL and DGLAP dynamics. 
We show  that using our  DLLA result HERA data 
can be described in the range  $ Q^2 \ge 150 \,GeV^2$ and all 
interval of $x$. Moreover, we compare our results with the predictions of 
Navelet {\it et al.} and with the predictions obtained using DGLAP 
evolution equations in the small-$x$ limit.
This letter is organized as follows. In Section 2 the QCD dipole picture is 
presented. In  Section 3 we obtain the proton structure function in this 
approach and its DLLA limit. In section 4 we apply our result to $F_2$ HERA 
data and present 
our  conclusions.

\section{QCD dipole picture}
 
 In this section we describe the basic ideas of the QCD dipole picture in 
the onium-onium scattering. Let ${\cal{A}}$ be the scattering amplitude 
normalized according to

\begin{eqnarray}
\frac{d\sigma}{dt}=\frac{1}{4\pi} |{\cal{A}}|^2\,\,\,.
\end{eqnarray}

 The scattering amplitude is given by 

\begin{eqnarray}
{\cal{A}} = -i\int d^2x_{1}d^2x_{2}\int dz_1dz_2 
\Phi^{(0)}(\underline{x}_{1},z_1)\,\Phi^{(0)}(\underline{x}_{2},z_2) {\cal{F}}(\underline{x}_{1}, \underline{x}_{2})\,\,\,,
\label{ampli1}
\end{eqnarray}
where $\Phi^{(0)}(\underline{x}_{i},z_i)$ is the squared wave function of 
the quark-antiquark part of the  onium wavefunction,
$\underline{x}_{i}$ being the transverse size of the quark-antiquark pair 
and $z_i$ the longitudinal momentum fraction of 
the antiquark. In lowest order  ${\cal{F}}$ is the elementary 
dipole-dipole cross-section    $\sigma_{DD}$.

In the large $N_c$ limit and in the leading-logarithmic-approximation the 
radiative corrections are generated by emission of gluons with strongly 
ordered longitudinal momenta fractions $z_i >> z_{i+1}$. The onium wave 
function with $n$ soft gluons  can be calculated using perturbative QCD. In 
the Coulomb gauge the soft radiation can be viewed as a cascade of colour 
dipoles emanating from the initial $q\overline{q}$ dipole, since each gluon 
acts like a quark-antiquark pair. Following \cite{mueller, mueller1}, we 
define the dipole density $n(Y,\underline{x},\underline{r})$  such that 
\begin{eqnarray}
N(Y,\underline{r})=\int dz_1 \int d^2 \underline{x} \,\Phi^{(0)}(\underline{x},z_1)\,n(Y,\underline{x},\underline{r})
\end{eqnarray}
is the number of dipoles of transverse size $\underline{r}$ with the smallest 
light-cone momentum in the pair   greater than or equal to $e^{-Y}p_+$, 
where $p_+$ is the light-cone momentum of the onium. The whole dipole cascade can be constructed from a 
repeated action of a kernel ${\cal{K}}$ on the initial density 
$n_o(\underline{x},\underline{r})$ through the dipole 
evolution equation
\begin{eqnarray}
n(Y,\underline{x},\underline{r}) = n_o(\underline{x},\underline{r}) +
\int_0^Y dy \int_0^{\infty} d\underline{s}\, {\cal{K}}\,(\underline{r},\underline{s}) n(y,\underline{x},\underline{s})\,\,\,.
\label{bfkl}
\end{eqnarray}

The evolution kernel $\cal{K}$ is calculated in perturbative QCD. For 
fixed $\alpha_s$ and in the limit $N_c \rightarrow \infty$ the kernel has 
the same spectrum as the BFKL kernel. Consequently, the two approaches lead to
the same phenomenological results for inclusive observables.
The solution of (\ref{bfkl}) is given by \cite{mueller, mueller1}
\begin{eqnarray}
n(Y,\underline{x},\underline{r}) = \frac{1}{2} \frac{x}{r} \frac{exp[(\alpha_P-1)Y]}{\sqrt{7 \alpha C_F \zeta (3)Y}}
exp\left(-\frac{\pi ln^2(x/r)}{28 \alpha C_F \zeta (3)Y}\right)\,\,\,,
\label{n}
\end{eqnarray}
where $\alpha_P-1 = (8 \alpha C_F/ \pi)ln2$.

The onium-onium scattering amplitude in the leading-logarithmic approximation 
will be written as in (\ref{ampli1}), but where $\cal{F}$ is now given by 
\begin{eqnarray}
{\cal{F}} = \int \frac{d^2\underline{r}}{\underline{r}}\frac{d^2\underline{s}}{\underline{s}}n(Y/2,\underline{x}_1,\underline{r}) n(Y/2,\underline{x}_2,\underline{s})\, \sigma_{DD} \,\,.
\label{ampli2}
\end{eqnarray}
Consequently, the cross section grows rapidly with the energy because the 
number of dipoles in the light cone wave function grows rapidly with the 
energy.
This result is valid in the kinematical region where $Y$ is not  very large. 
At large $Y$ the cross section breaks down due to the diffusion to large 
distances, determined by the last exponential factor in (\ref{n}), and  due to 
the unitarity constraint. Therefore new corrections should become important 
and modify the BFKL behaviour \cite{fad}.

The result (\ref{n}) was obtained for a process with only one scale, the 
onium radius. In processes where two scales are present, for example the 
eletron-proton deep inelastic scattering, this result is affected by 
non-perturbative  contributions \cite{bartels}. Therefore, the application 
of BFKL approach at $ep$ scattering must be made with caution.

% The acquisition of  results via BFKL approach in agreement with $F_2$ data no %signify that the small $x$ dynamics is the BFKL. Since the non-perturbative %contribution is unknown, the results  have free parameters that can be adjusted %of accord with the data.   

\section{Structure function in the QCD dipole picture}

Our goal in this section is to obtain the proton structure functions
\begin{eqnarray}
F_{L,T}(x,Q^2) = \frac{Q^2}{4 \pi \alpha_{e.m.}}\,\sigma^{\gamma^*\, p}_{L,T}\,\,\,
\end{eqnarray}
using the QCD dipole picture. In order to do so we must make the assumption that the proton can be approximately described by onium configurations. Basically, we make use of the assumption
\begin{eqnarray}
\sigma^{\gamma^*\, p}_{L,T} = \sigma^{\gamma^* \,onium}_{L,T} \times {\cal{P}}\,\,\,,
\end{eqnarray}
where ${\cal{P}}$ is the probability of finding an onium in the proton.  In order to obtain the $\sigma^{\gamma^* \,onium}_{tot}$ we will follow \cite{navelet}, where the $k_T$ factorization \cite{catani} was used in the context of the QCD dipole model.  We have that 
\begin{eqnarray}
\sigma^{\gamma^* \,onium}_{L,T} =  \int d^2 \underline{r}\, dz \,\Phi^{(0)}(\underline{r},z)\, \sigma^{\gamma^* \, dipole}(x,Q^2,\underline{r})\,\,\,,
\label{sig1}
\end{eqnarray}
where the  $\gamma^* - dipole$ cross section reads
\begin{eqnarray}
Q^2\sigma^{\gamma^* \, dipole}(x,Q^2,\underline{r}) = \int d^2\underline{k} \int_0^1 \frac{dz}{z} \,\hat{\sigma}_{\gamma^*\,g}(\frac{x}{z}, \frac{\underline{k}^2}{Q^2})\,{\cal{G}}(z,\underline{k}, \underline{r})\,\,\,.
\label{sig2}
\end{eqnarray}
In (\ref{sig2})  $\hat{\sigma }_{ \gamma^* \, g}$ is the $\gamma^* g \rightarrow q \overline{q}$ Born cross section and  ${\cal{G}}$  is the non-integrated gluon distribution function. The relation between this 
 function and  the  dipole density is expressed by
\begin{eqnarray}
\underline{k}^2 \,{\cal{G}}(z,\underline{k}, \underline{r}) = \int \frac{\underline{s}^2}{\underline{s}} \int_0^1 \frac{dz\prime}{z\prime} n(z\prime ,\underline{r}, \underline{s}) \, \hat{\sigma}_{\gamma^* \,d}(\frac{z}{z\prime}, \underline{s}^2 \underline{k}^2)\,\,\,,
\label{sig3}
\end{eqnarray}
where $n(z\prime, \underline{r}, \underline{s})$ is the density of dipoles of transverse size $\underline{s}$ with the smallest light-cone momentum in the pair equal to $z\prime p_+$ in a dipole of transverse size $\underline{r}$, of total momentum $p_+$.  This is 
given by the solution (\ref{n}).

After some considerations Navelet {\it et al.} obtain  
\begin{eqnarray}
Q^2\sigma^{\gamma^* \, dipole}(x,Q^2,\underline{r}) = 4 \pi^2 \alpha_{e.m.} \frac{2 \alpha N_c}{\pi} \int \frac{d \gamma}{2 \pi i} h_{L,T}(\gamma) \frac{v(\gamma)}{\gamma}(\underline{r}^2Q^2)^{\gamma} e^{[ \frac{\alpha N_c}{\pi} \chi (\gamma) ln\frac{1}{x}]}\,,  \nonumber\\
\label{sig4}
\end{eqnarray}
where $ \frac{\alpha N_c}{\pi} \chi (\gamma)$ is the BFKL spectral function (for more details see \cite{navelet}).

The $\gamma^* \, onium$  cross section is obtained using (\ref{sig4}) in (\ref{sig1}). The result depends on the   squared wave function $\Phi^{(0)}$ 
of the onium state, which cannot be computed perturbatively. Consequently, an assumption must be made. In \cite{navelet} this dependence is eliminated by averaging over the wave function  of transverse size
\begin{eqnarray}
\int dz d^2\underline{r} (\underline{r}^2)^{\gamma} \Phi^{(0)}(\underline{r},z) = (M^2)^{-\gamma}\,\,\,,
\end{eqnarray}
where $M^2$ is a scale which is assumed to be perturbative.
Therefore the proton structure functions reads
\begin{eqnarray}
F_{L,T}(x,Q^2) = \frac{2 \alpha N_c}{\pi} \int \frac{d \gamma}{2 \pi i} h_{L,T}(\gamma) \frac{v(\gamma)}{\gamma}(\frac{Q^2}{M^2})^{\gamma} e^{[ \frac{\alpha N_c}{\pi} \chi (\gamma) ln\frac{1}{x}]} {\cal{P}}(\gamma, M^2) \,\,\,,
\end{eqnarray}
where $ {\cal{P}}(\gamma, M^2)$ is the Mellin-transformed probability of finding an onium of transverse mass $M^2$ in the proton.  Using an adequate choice for this probability (see ref. \cite{navelet}) the proton structure function can be written as 
\begin{eqnarray}
F_{L,T}(x,Q^2) = \frac{2 \alpha N_c}{\pi} \int \frac{d \gamma}{2 \pi i} h_{L,T}(\gamma) \frac{v(\gamma)}{\gamma}(\frac{Q^2}{Q_0^2})^{\gamma} e^{[ \frac{\alpha N_c}{\pi} \chi (\gamma) ln\frac{1}{x}]} {\cal{P}}(\gamma) \,\,\,.
\label{f2}
\end{eqnarray}
  
The expression (\ref{f2}) can be  evaluated using the steepest descent method. 
The saddle point is given by
\begin{eqnarray}
\chi^{\prime}(\gamma_s) = -  \frac{ln \frac{Q^2}{Q_0^2}}{\frac{\alpha N_c}{\pi} ln \frac{1}{x}}\,\,.
\label{sad}
\end{eqnarray}

Using in the expression (\ref{sad}) the expansion of the BFKL kernel near $\gamma = \frac{1}{2}$, we get
\begin{eqnarray}
\gamma_s = \frac{1}{2} \left( 1 -\frac{ \frac{Q}{Q_0}}{\frac{\alpha N_c}{\pi} 7 \zeta (3) ln \frac{1}{x}} \right)\,\,.
\end{eqnarray}
Consequently 
\begin{eqnarray}
F_2(x,Q^2)  & = & F_T(x,Q^2) + F_L(x,Q^2) \nonumber \\
            & = & C \, a^{\frac{1}{2}} \frac{Q}{Q_0} e^{[(\frac{4\alpha N_c ln 2}{\pi})ln \frac{1}{x} - \frac{a}{2} ln^2 \frac{Q}{Q_0}]}\,\,\,,
\label{f2nav}
\end{eqnarray}
where
\begin{eqnarray} 
a = \left(\frac{1}{\frac{\alpha N_c}{\pi}7\zeta (3) ln \frac{1}{x}}\right)\,\,\,.
\end{eqnarray}
The parameters $C$, $Q_0$ and $\alpha$ are determined by the fit. Using 
$C = 0.077$, $Q_0 = 0.627\, GeV$ and $\alpha \approx 0.11$, Navelet {\it et al.}  obtained that the expression (\ref{f2nav}) fits the  HERA data \cite{h1}  in the region $Q^2 \le 150 \,GeV^2$.

In this letter we analize the behaviour of $F_2$ obtained by expression 
(\ref{f2}) in the double leading logarithmic approximation (DLLA). 
This limit is common to both DGLAP and BFKL dynamics, {\it i.e.} it represents 
the transition region between the dynamics. Therefore, before the determination 
of the region  where the BFKL dynamics (BFKL Pomeron) is valid, we must 
determine clearly the region where the DLLA limit is valid.  In this limit 
$\chi(\gamma) = \frac{1}{\gamma}$. Using this limit in (\ref{sad}) 
we get that  the  saddle point is at
\begin{eqnarray}
\gamma_s = \sqrt{\frac{\frac{\alpha N_c}{\pi}ln \frac{1}{x}}{ln \frac{Q^2}{Q_0^2}}}\,\,\,.
\label{cela}
\end{eqnarray}
Consequently, we get
\begin{eqnarray}
F_2^{DLLA}(x,Q^2) = \frac{2 \alpha N_c}{\pi} \, {\cal{C}} \, \frac{(ln \frac{Q^2}{Q_0^2})^{\frac{1}{4}}}{(\frac{\alpha N_c}{\pi} ln \frac{1}{x})^{\frac{3}{4}}} \,
e^{\left[2\sqrt{\frac{ \alpha N_c}{\pi} ln \frac{1}{x} ln \frac{Q^2}{Q_0^2}}\right]}\,\,.
\label{f2vic}
\end{eqnarray}
The result (\ref{f2vic}) reproduces the behaviour of double leading 
logarithmic approximation. As this result was obtained considering the dipole
model, then $\alpha_s$ is fixed. The parameters ${\cal{C}}$, $\alpha$ and $Q_0$  
should be taken from the fit.  In the next section we compare  this result with  HERA data.  

\section{Results and Conclusions}

In this section we  compare the expression (\ref{f2vic}) with the  recent H1 data \cite{h1}. In order to test the accuracy of the $F_2$ parameterization obtained in formula  (\ref{f2vic}), a fit of H1 data  has been performed. The parameters obtained were
\begin{eqnarray}
{\cal{C}} =  0.0035\,,\,\,\,\,Q_0=0.45\,\,\mbox{and}\,\,\alpha = 0.19\,\,.
\end{eqnarray}

In the figures (\ref{fig1}) and  (\ref{fig2})  we present  our results  at 
different $Q^2$ and $x$. The predictions of Navelet {\it et al.} (dashed curve) 
are also presented. While the expression (\ref{f2nav}) describes the data  
in region $Q^2\le150\,GeV^2$, we can see that the expression (\ref{f2vic}) 
(solid curve) describes H1 data in  kinematical region $ Q^2\,>\,150\, GeV^2$.   
Therefore the HERA data are  described by the DLLA expression  in the high $Q^2$ region. 
Moreover, we can conclude that there is a transition  between the 
BFKL behaviour  and the DLLA behaviour  in the region $Q^2 \approx 150 \, GeV^2$. This 
could  be the first evidence of the BFKL behaviour in $F_2$. However, this 
conclusion is not strong since that (\ref{f2nav}) and (\ref{f2vic}) were  
obtained in the leading-logarithmic-approximation, where $\alpha_s$ is fixed.
Moreover, the  parametrizations obtained using the DGLAP evolution 
equation describes the HERA data \cite{grv95}. 

The program of calculating the next-to-leading corrections to the BFKL equation is not still concluded \cite{fad}. However, some results may be anticipated.
For instance, the NLO BFKL equation must have as limit the DLLA limit in the region where
$\alpha_s\,log\,\frac{1}{x}\,log\,Q^2\,\approx\,1$. 
This limit  is common to both DGLAP and BFKL dynamics. As the  NLO corrections
to the DGLAP evolution equations are known, the DLLA limit with $\alpha_s$ running is well established. The DLLA limit obtained considering the DGLAP evolution equation 
 was largely discussed by Ball and Forte \cite{ball}. 
Consequently we can estimate the importance of $\alpha_s$ running in our 
result.
In figure (\ref{fig2}) we compare our results with the  DLLA predictions  
with $\alpha_s$ running (dot-dashed curve), obtained using DGLAP evolution equations in the small-$x$ limit. 
In this case the DLLA limit can describe one more large kinematical region.
The region $Q^2 \le 5 \,GeV^2$ is not described by DLLA $\alpha_s$ running.
Consequently, the kinematical region where the BFKL dynamics may be present is restricted 
to the low $Q^2$ region. 

Our 
results are strongly dependent on  free parameters, since   there is great freedom in the choice of 
parameters used in both BFKL and DLLA predictions.  This   comes from  
theoretical uncertainties, for example,   the determination where the 
pQCD is valid ({\it i.e.} the $Q_0$ value). However,
our qualitative result agrees  with the conclusion obtained
by Mueller \cite{ope}, that demonstrated that the BFKL diffusion
leads to the breakdown of the OPE at small-$x$. 
Using the bounds obtained by Mueller we expect that the BFKL evolution can be 
visible in a limited range of $x$ at $Q^2 \rightarrow Q_0^2$.
Moreover, our result agrees with the conclusion of Ayala {\it et al.} \cite{ayala}, where 
the shadowing corrections to $F_2$ structure function were considered.
In this case the anomalous dimension is
modified and the BFKL behaviour only can be visible in the region of 
low $Q^2$.

In this paper we calculate the DLLA limit of BFKL in the dipole picture.  The determination of this limit is very important, since it is common to BFKL and DGLAP dynamics. Therefore, before the determination of the region where the BFKL dynamics is valid, we must determinate clearly the region where the DLLA limit is valid. We demonstrate that in the leading-logarithmic approximation ($\alpha_s$ fixed) a transition  region between the BFKL dynamics and the  
DLLA limit  can be obtained in the region of $Q^2 \approx \,150\,GeV^2$.
We compare this result with the  DLLA predictions obtained with 
$\alpha_s$ running. In this case a transition region is obtained at low $Q^2$
($ \le 5\,GeV^2$).  
This demonstrates the importance of the next-to-leading order corrections
to the BFKL dynamics.
  Our conclusion is  that the
 $F_2$ structure function is not  the best observable in the 
determination of the dynamics.  From the inclusive measurements of $F_2$ it seems improbable  to 
draw any conclusion based on the presently available data. It is theoretically 
questionable whether it will be  possible as long as one  considers only one 
observable.  The better observables for  determination of the dynamics are the 
ones associated with processes where only one scale is present, since  
in these processes it  is inambigously   possible to isolate the effects of 
the BFKL behaviour.

\section*{Acknowledgments}

This work was partially financed by CNPq, BRAZIL.

\newpage
\section*{Figure Captions}

\begin{figure}
%\begin{tabular}{c c}
\centerline{\psfig{file=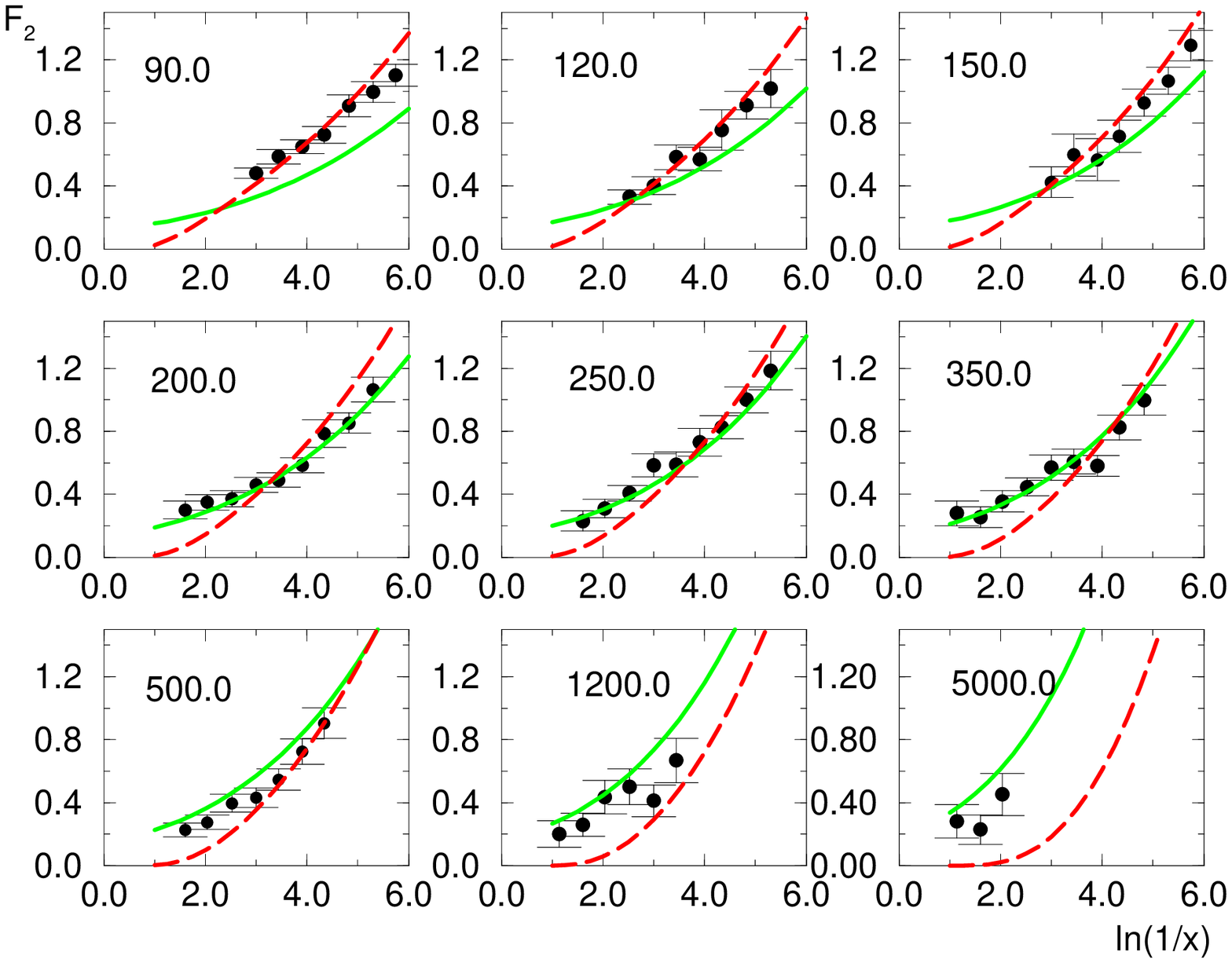,width=150mm}} 
%\end{tabular}
\caption{}
\label{fig1}
\end{figure}

\vspace{1.0cm}
Fig. \ref{fig1}: Behavior of proton structure function predicted by BFKL (\ref{f2nav}) (dashed curve) and  DLLA (\ref{f2vic})  (solid curve). Data of  H1 \cite{h1}. See text.

\vspace{1.0cm}

\begin{figure}
%\begin{tabular}{c c}
\centerline{\psfig{file=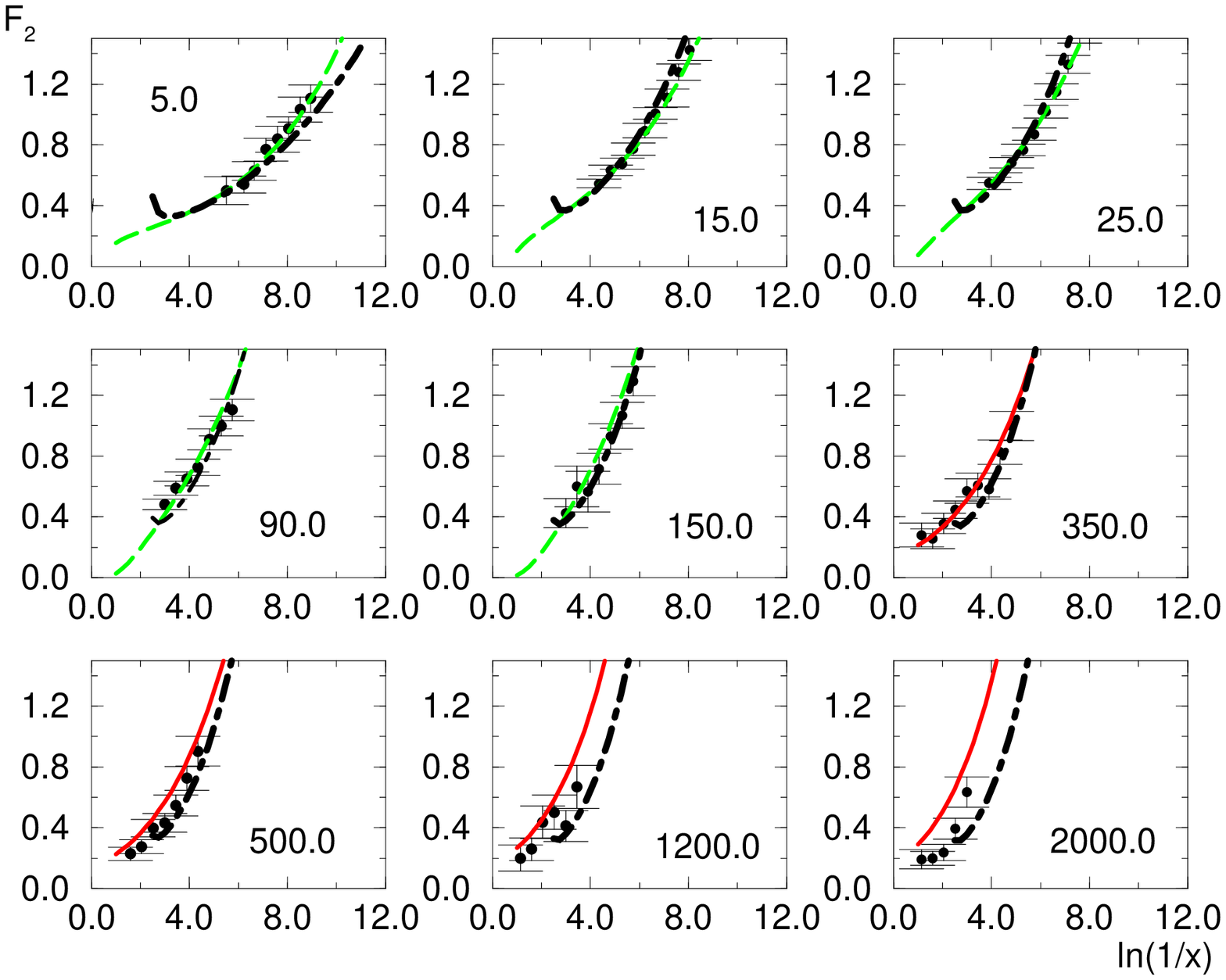,width=150mm}} 
%\end{tabular}
\caption{}
\label{fig2}
\end{figure}

Fig. \ref{fig2}: Behavior of proton structure function predicted by BFKL (\ref{f2nav}) (dashed curve), DLLA (\ref{f2vic})  (solid curve) and DLLA with running coupling constant (dot-dashed curve). Data of  H1 \cite{h1}. See text.

\newpage

%\begin{figure}[t]
%\begin{tabular}{c c}
%\psfig{file=fig5.eps,width=70mm} & \psfig{file=fig6.eps,width=70mm} %\end{tabular}
%\caption{}
%\label{fig2}
%\end{figure}

%\begin{figure}[t]
%\begin{tabular}{c c}
%\psfig{file=fig3.eps,width=70mm} & \psfig{file=fig4.eps,width=70mm} %\end{tabular}
%\caption{}
%\label{fig3}
%\end{figure}

%\begin{figure}
%\centerline{\psfig{file=fig1.eps,width=150mm}}
%\caption{}
%\label{fexp2}
%\end{figure}

%\begin{figure}
%\centerline{\psfig{file=gscren5.eps,width=150mm}}
%\caption{}
%\label{gscren}
%\end{figure}

\end{document}